\PassOptionsToPackage{table}{xcolor}
\documentclass[letterpaper]{article} 
\usepackage[preprint]{aaai2027} 
\usepackage[hyphens]{url}  
\usepackage{graphicx} 
\usepackage{float} 
\usepackage{natbib}  
\usepackage{caption} 
\usepackage{amsmath}
\usepackage{amssymb}
\usepackage{booktabs}
\usepackage{array}

\definecolor{GRBControl}{RGB}{242,242,242}
\definecolor{GRBPhenotype}{RGB}{232,244,252}
\definecolor{GRBLLM}{RGB}{244,235,250}
\definecolor{GRBGraph}{RGB}{232,246,238}
\definecolor{GRBProbe}{RGB}{255,243,224}
\definecolor{GRBHeatCyan0}{HTML}{F7FCFD}
\definecolor{GRBHeatCyan1}{HTML}{E4F4F6}
\definecolor{GRBHeatCyan2}{HTML}{CBE9EC}
\definecolor{GRBHeatCyan3}{HTML}{A9DADF}
\definecolor{GRBHeatCyan4}{HTML}{83C6CE}
\definecolor{GRBHeatCyan5}{HTML}{5AAFB9}
\definecolor{GRBHeatCyan6}{HTML}{2F8F9B}

\title{GraphRareBench: An Auditable Graph-Evidence Benchmark for Phenotype-Driven Rare-Disease Diagnosis}

\author{
Guiling Guo\textsuperscript{\rm 1},
Jia Yang\textsuperscript{\rm 2},
Jiahao Xu\textsuperscript{\rm 3},
Shuyuan Zheng\textsuperscript{\rm 3},\\
Zhonghai Sun\textsuperscript{\rm 4},
Qiyuan Li\textsuperscript{\rm 3,1}
}

\affiliations{
\textsuperscript{\rm 1}National Institute for Data Science in Health and Medicine, Xiamen University\\
\textsuperscript{\rm 2}Institute of Artificial Intelligence, Xiamen University\\
\textsuperscript{\rm 3}School of Medicine, Xiamen University\\
\textsuperscript{\rm 4}The First Affiliated Hospital of Xiamen University
}

\begin{document}

\maketitle

\begin{abstract}
Phenotype-driven diagnostic benchmarks usually report the rank of the reference disease, but they rarely reveal which plausible alternatives are ranked above it or what evidence a tool-using model examines before making its decision. We introduce GraphRareBench, a provenance-preserving benchmark containing 2,365 ontology-derived cases and 18,093 target--confounder pairs. Each case includes a coarsened HPO query, a fixed candidate pool, graph-defined hard confounders, and source-linked evidence records. On the 237-case gene-component-disjoint test split, supervised rankers using a shared 21-feature interface achieved MRRs of 0.640–0.740 and case-averaged target-over-confounder accuracies of 0.898–0.916. Agents instantiated with Agents-A1 and DeepSeek-V4-Flash achieved MRRs of 0.746 and 0.718, respectively. Their paired MRR difference was not statistically significant, whereas their target-evidence coverage differed by 0.561. Together with the observation that 22.1\%–43.7\% of selected Hit@10 successes still ranked at least one graph-defined hard confounder above the target, these results indicate that full-pool retrieval, hard-confounder discrimination, and observable evidence access capture complementary aspects of model behavior. GraphRareBench therefore provides a foundation for more transparent and evidence-aware evaluation of phenotype-driven diagnostic systems. Code and dataset are avaliable at https://github.com/GUI0609/GraphRareBench.
\end{abstract}

\begin{figure}[H]
\centering
\captionsetup{font=small}
\includegraphics[width=0.93\textwidth,height=0.39\textheight,keepaspectratio]{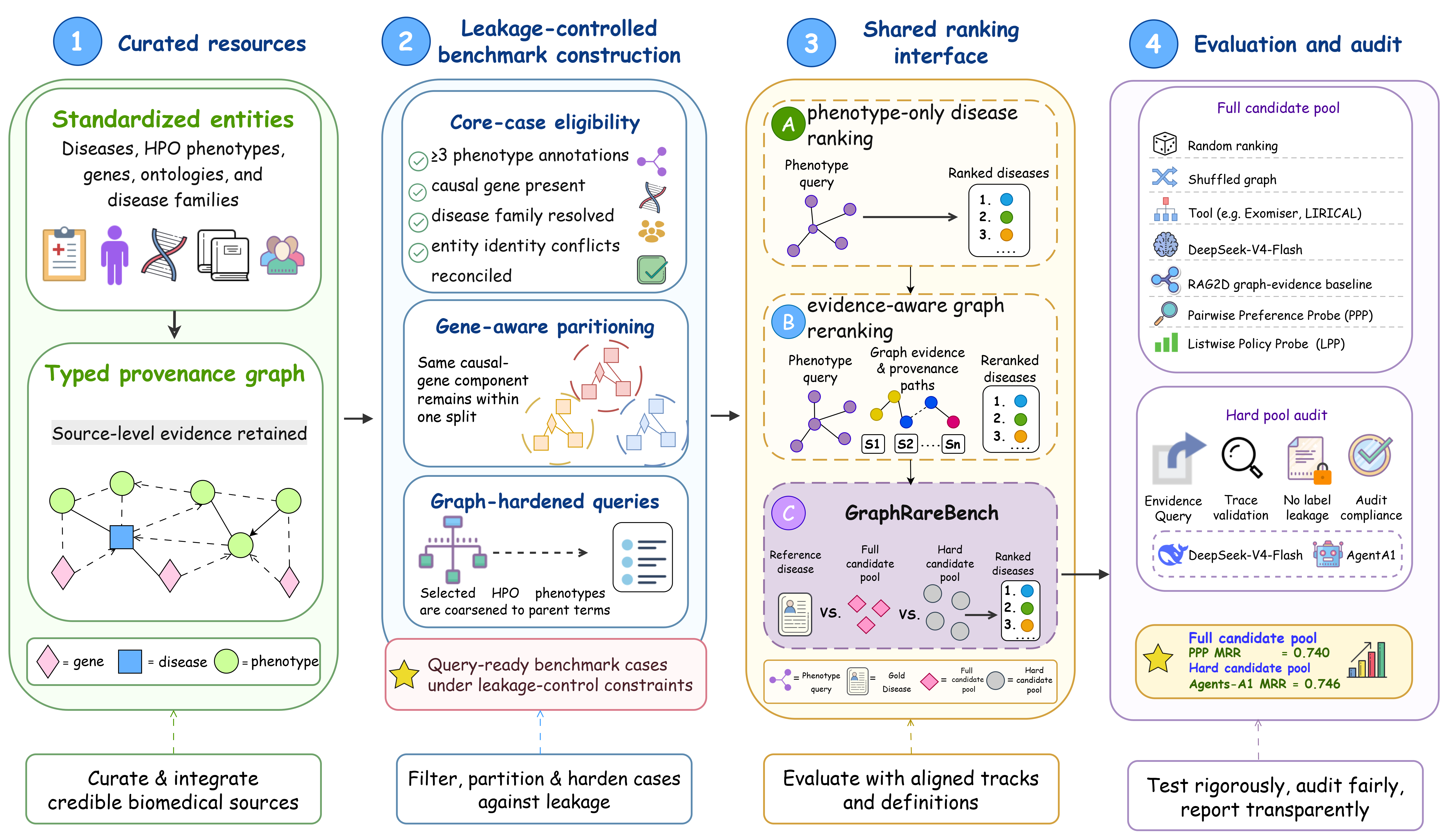}
\caption{
GraphRareBench converts curated rare-disease resources into ontology-derived
ranking cases with fixed candidate pools, graph-defined hard alternatives, and
source-linked evidence records. Evaluated systems receive the declared
query, candidate pool, and evidence view; evaluator-only labels support
target-over-confounder, mechanism-sliced, and tool-trace audits.
Icons are from Icons8.com.
}
\label{fig:overview}
\end{figure}
\section{Introduction}

Rare diseases are individually uncommon but collectively impose a substantial diagnostic burden. Many patients undergo prolonged diagnostic odysseys because early clinical presentations are heterogeneous, incomplete, and shared across multiple disorders. Although molecular testing has become central to confirming many genetic rare diseases, phenotype-driven reasoning remains essential for generating diagnostic hypotheses, selecting tests, prioritizing variants, guiding referrals, and constructing differential diagnoses \citep{manickam2021acmg}. Computational diagnostic systems must therefore address a fundamental clinical task: ranking candidate rare diseases from a compact and often incomplete phenotype description.

This task is inherently difficult because a small set of observed phenotypes may support several closely related diseases. Candidate diseases may share HPO terms, ontology ancestors, disease families, semantic neighborhoods, gene families, causal-gene components, or even the same causal gene. Standard top-$k$ metrics capture whether the target disease appears near the top of the list, but not whether a clinically plausible alternative is ranked above it. A model may therefore achieve a favorable Hit@$k$ score while still failing at the central differential-diagnosis problem. Meaningful evaluation must measure both target retrieval and the model's ability to distinguish the target from closely related alternatives.

This limitation becomes more consequential for large language model (LLM)-based diagnostic systems. LLMs can generate fluent and medically plausible rankings, yet the final output alone does not reveal whether the ranking is supported by case-specific evidence or by broad disease associations and prior knowledge. Moreover, final-ranking metrics alone may fail to reveal substantial differences in the evidence-seeking trajectories of tool-using systems. Evaluating these systems therefore requires more than a reference-disease rank: the candidate set must contain difficult alternatives, the evidence available for each candidate must remain traceable, and the model's evidence-access behavior must be auditable.

We introduce GraphRareBench, a provenance-preserving benchmark for phenotype-driven rare-disease ranking. Each case contains a coarsened HPO query, a fixed full candidate pool, a graph-defined subset of hard confounders, and evidence records linked to their original sources. Target and confounder annotations are retained by the evaluator, while evaluated methods receive only the declared query, candidate set, and permitted evidence interface. This design supports two complementary evaluations: full-pool ranking measures whether a method retrieves the target disease among a broad set of candidates, whereas a tool-mediated hard-pool audit examines how an agent distinguishes the target from its closest alternatives and which evidence it accesses before producing the final ranking.

\paragraph{Contributions.}

Our contributions are threefold: (i) we introduce GraphRareBench, comprising 2,365 ontology-derived cases and 18,093 target--confounder pairs, with coarsened HPO queries, fixed candidate pools, gene-component-aware data partitions, seven graph-defined confounder mechanisms, and source-linked evidence records; (ii) we provide target-over-confounder metrics and mechanism-specific analyses that directly evaluate whether a model can distinguish the target disease from closely related alternatives; and (iii) we establish a unified evaluation framework for both final ranking performance and observable evidence-access behavior, covering phenotype-driven tools, prompted LLMs, supervised graph-evidence rankers, and tool-using agents.

\section{Related Work}

\paragraph{Rare-disease resources and phenotype-driven tools.}
Phenotype-driven rare-disease diagnosis relies on standardized vocabularies, disease ontologies, molecular databases, and knowledge graphs that connect clinical phenotypes with diseases, genes, variants, and diagnostic panels. HPO, Orphanet, OMIM, and Mondo provide complementary representations of phenotypes and rare diseases, while GenCC, ClinGen, ClinVar, PanelApp, and KGRD contribute gene--disease validity, variant interpretation, diagnostic-panel information, and graph-structured biomedical evidence \citep{gargano2024hpo,rath2012orphanet,amberger2015omim,vasilevsky2026mondo,distefano2022gencc,andersen2025clingen,landrum2025clinvar,martin2019panelapp,guo2026kgrd}. Phenotype-driven systems such as Phenomizer, Exomiser, LIRICAL, Phen2Gene, and PhenoBrain use these resources to prioritize candidate diseases, genes, or variants from patient phenotypes \citep{kohler2009clinical,smedley2015exomiser,robinson2020lirical,zhao2020phen2gene,mao2025phenobrain}.

\paragraph{Tool-using medical agents and trace evaluation.}
Tool-using medical agents turn rare-disease diagnosis into a sequence of information-processing and evidence-integration steps rather than a single free-form prediction. RareAgents combines multidisciplinary-team coordination, long-term memory, and specialized phenotype and treatment-oriented tools, allowing different physician agents to retrieve similar cases, query diagnostic systems, exchange findings, and synthesize a final report \citep{chen2026rareagents}. DeepRare extends this pattern with a central coordinating agent and more than 40 specialized tools and knowledge sources for phenotype normalization, disease normalization, case retrieval, phenotype and genotype analysis, literature search, and evidence-linked differential diagnosis \citep{zhao2026deeprare}. KGRD adds knowledge-graph-guided gene and phenotype inference, patient-level case retrieval, multidisciplinary specialist deliberation, and a verifier that checks candidate support across multiple evidence channels \citep{guo2026kgrd}. These systems therefore expose richer intermediate artifacts than one-shot LLMs, including retrieved cases, tool outputs, agent discussions, graph paths, and source-linked evidence. Such artifacts make it possible to ask not only whether the final diagnosis is correct, but also which evidence channels were selected, whether alternatives were examined, and how the final ranking was assembled.

\paragraph{Rare-disease evaluation benchmarks.}
Phenopacket Store and PhEval provide standardized case representations and reproducible evaluation protocols, while RareBench, RareArena, MIMIC-RD, UDN-derived studies, RareSyn, and SHEPHERD extend benchmarking across clinical cases, language-model diagnosis, and graph-based systems \citep{danis2025phenopackets,bridges2025pheval,chen2024rarebench,chen2026rarearena,eizaldin2025mimicrd,shyr2025udn,wang2025raresyn,alsentzer2025shepherd}. Collectively, these resources have strengthened the evaluation of target-disease retrieval, but their primary outcomes remain rank-based measures such as reciprocal rank, top-\(k\) retrieval, or candidate-ranking accuracy. Large-scale comparisons further suggest that, for phenotype-only diagnosis, LLMs remain less reliable than established rare-disease decision-support tools \citep{reese2026llmbenchmark}. In parallel, general agent benchmarks such as AgentBench, ToolSandbox, and \(\tau\)-bench emphasize that tool-use validity, intermediate actions, and task completion should be evaluated in addition to final answers \citep{liu2024agentbench,lu2025toolsandbox,yao2024taubench}. However, existing medical and agent benchmarks rarely connect observable action traces to a fixed differential-diagnosis candidate set containing explicitly labeled hard confounders and source-traceable evidence. GraphRareBench addresses this gap by fixing the candidate pool, identifying hard alternatives through graph relations, and jointly auditing the final ranking and the evidence requested during tool use.

\section{Task Formulation}
\label{sec:task-formulation}

GraphRareBench formulates phenotype-driven rare-disease ranking as a closed candidate-set task. Each case \(i\) is derived from an ontology-linked disease profile and contains one target disease together with a set of candidate alternatives. The input visible to an evaluated method is
\[
X_i=(Q_i,C_i,E_i),
\]
where \(Q_i\) is the coarsened phenotype query, \(C_i\) is the candidate set used in the current evaluation, and \(E_i\) is the evidence view permitted by the corresponding interface. The target disease \(g_i\in C_i\) is hidden from the method and used only for evaluator-side scoring.

GraphRareBench provides two candidate-set configurations. The full candidate pool is
\[
C_i^{\mathrm{full}}=\{g_i\}\cup H_i\cup A_i,
\]
where \(H_i\) contains graph-defined hard confounders and \(A_i\) contains additional candidate diseases. This pool evaluates whether a method can retrieve the target from a broader set of alternatives. The separate tool-mediated audit uses the hard candidate pool
\[
C_i^{\mathrm{hard}}=\{g_i\}\cup H_i,
\]
which focuses on distinguishing the target from its closest graph-defined alternatives.

Each evaluated method must return a complete ranking of all diseases in the active candidate set \(C_i\). Target identities, hard-confounder annotations, and confounder-reason labels are never exposed through the evaluation interface. Let \(\mathrm{rank}_i(c)\) denote the one-indexed position of candidate \(c\) in the returned ranking.

\subsection{Metrics}

We report standard MRR and Hit@\(k\) for target retrieval. Let
\(\mathrm{rank}_i(c)\) denote the one-indexed rank of candidate \(c\).
For cases containing hard confounders, target-over-confounder accuracy is
defined as
\[
\begin{aligned}
\mathrm{ToC}_i
&=\frac{1}{|H_i|}
  \sum_{h\in H_i}
  \mathbb{I}[\mathrm{rank}_i(g_i)<\mathrm{rank}_i(h)],\\
\mathrm{ToC}_{\mathrm{case}}
&=\frac{1}{N_C}
  \sum_{i:\,|H_i|>0}\mathrm{ToC}_i,\\
\mathrm{ToC}_{\mathrm{pair}}
&=\frac{\sum_{i:\,|H_i|>0}|H_i|\mathrm{ToC}_i}
        {\sum_{i:\,|H_i|>0}|H_i|},
\end{aligned}
\]
where \(N_C\) is the number of cases with at least one hard confounder.
Thus, \(\mathrm{ToC}_{\mathrm{case}}\) weights cases equally, whereas
\(\mathrm{ToC}_{\mathrm{pair}}\) weights target--confounder pairs equally.
We additionally report \(\mathrm{AnyConfAbove}\), the proportion of cases
with \(\mathrm{ToC}_i<1\). Mechanism-specific analyses compute
\(\mathrm{ToC}_{\mathrm{pair}}\) after restricting comparisons to pairs
carrying the corresponding confounder label.

\section{GraphRareBench Construction}

\paragraph{Resource harmonization.}
GraphRareBench is built from normalized rare-disease records that link disease identities to HPO phenotype annotations, causal genes, ontology relations, disease-family metadata, and source-level evidence. Starting from 7,441 normalized disease records, we retained 2,371 diseases that satisfied the predefined requirements for phenotype coverage, gene annotation, family metadata, and identity resolution. A final label-leakage screen removed six additional diseases, resulting in 2,365 released cases. Each retained disease was required to have at least three disease-level HPO annotations. The resulting graph contains 10,754 phenotype nodes, 4,262 gene nodes, 212,901 typed modeling edges, and 515,774 source assertions. We additionally derive a smaller model-visible evidence layer that supports controlled evidence access and provenance auditing. 

\paragraph{Gene-aware partitioning.}
To reduce information leakage across genetically related diseases, GraphRareBench uses causal-gene annotations when constructing the train, validation, and test partitions. Diseases connected through the same causal-gene component are assigned to the same partition whenever gene annotations are available. This procedure produces 1,892 training cases, 236 validation cases, and 237 test cases. Candidate mining is performed only among diseases assigned to the same partition as the target. Evidence records associated with test diseases remain available through the declared benchmark interfaces. The resulting setting therefore evaluates whether supervised rankers generalize to target diseases from causal-gene components not observed during training, while preserving a shared evidence-access environment at test time.

\paragraph{Query abstraction and evidence preservation.}
For each eligible target disease \(g_i\), GraphRareBench constructs a model-visible phenotype query \(Q_i\) together with a source-traceable evidence record. Across the 2,365 released cases, the visible queries contain 9,423 HPO terms, with a median of four terms per case and a range of two to four. Of these terms, 9,208 (97.7\%) are coarsened ontology parents, whereas 215 (2.3\%) are retained directly. All mappings were validated against the registered HPO OBO release dated 2026-02-16. Each coarsened term is exactly one upward ontology step from its source phenotype, while each direct term has distance zero. No case contains duplicate visible HPO identifiers.

The coarsening procedure modifies only the phenotype query presented to the model. The original source phenotypes are preserved in the provenance records and may be accessed only through the evidence interfaces permitted in the corresponding evaluation setting. This design reduces direct phenotype matching while retaining the evidence required for controlled retrieval and audit.

\paragraph{Candidate-pool construction and annotations.}
Each case uses the two fixed candidate views defined above. Hard confounders
are mined deterministically within each partition using high HPO overlap,
SapBERT semantic neighbors \citep{liu2021self}, shared direct MONDO parents,
disease-family membership, shared causal-gene families, shared causal-gene
components, and exact causal-gene sharing. Because these mechanisms overlap,
each target--confounder pair may receive multiple labels. Additional non-hard
candidates are drawn from semantic, phenotype, graph-neighbor, and
deterministic random sources after labeled hard confounders are excluded.
Candidate order is independently permuted, and all evaluator-side labels
remain hidden at test time. The frozen release contains 18,093 labeled
target--confounder pairs; detailed frequencies, overlaps, thresholds, and
candidate-cap analyses are reported in the supplementary material.

\begin{figure}[t]
\centering
\includegraphics[width=0.5\textwidth]{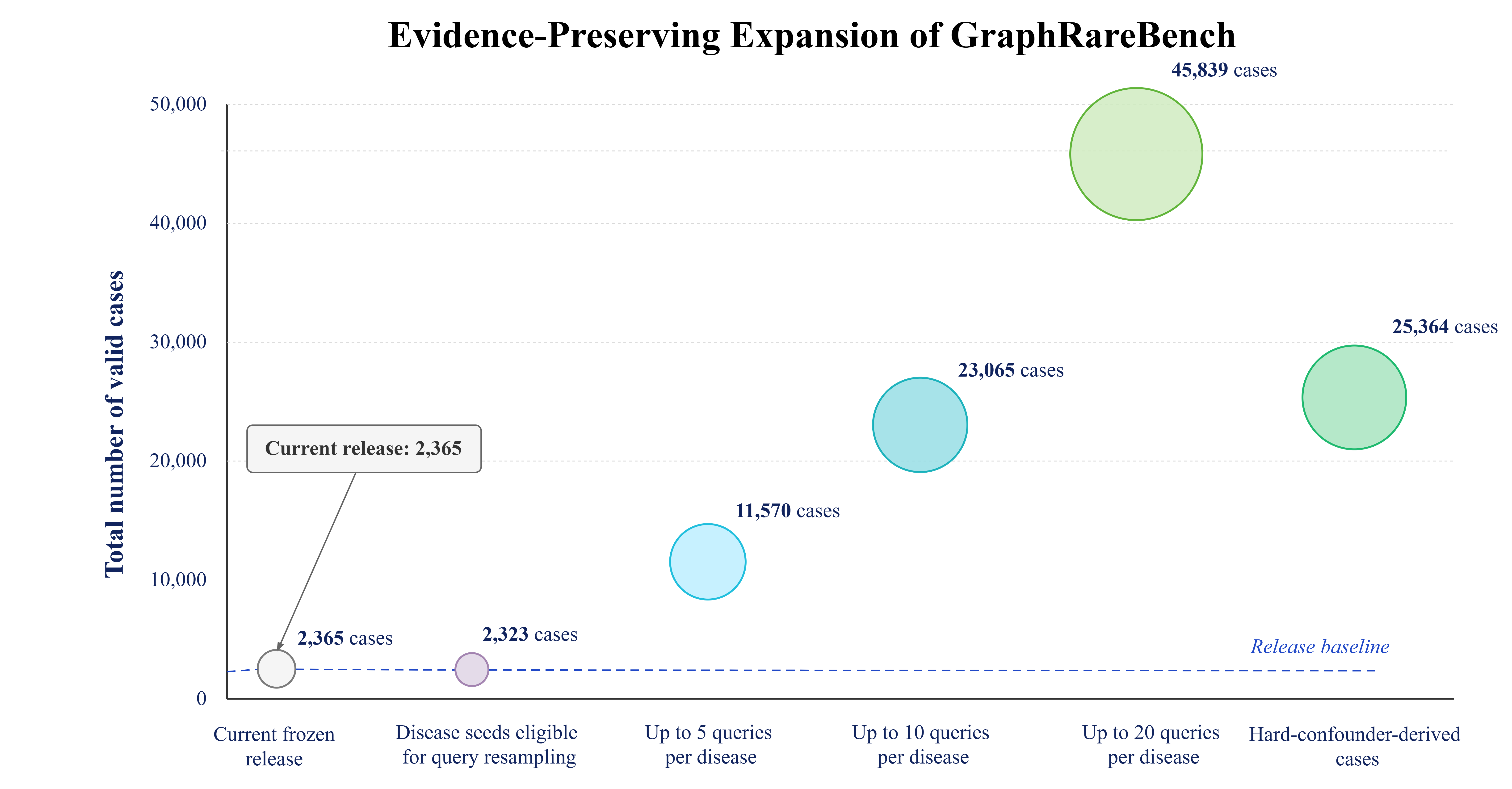}
\caption{Frozen-release scale and controlled expansion capacity. The current benchmark contains 2,365 fixed cases, of which 2,323 diseases are eligible for additional query construction. At query caps of 5, 10, and 20 per eligible disease, deterministic validity and deduplication filters yield 11,570, 23,065 and 45,839 disease-query cases, respectively. The hard-confounder-derived route yields 25,364 distinct disease-query cases.}
\label{fig:release-scale}
\end{figure}
\section{Experiments}
\label{sec:experiments}

We organize the experiments around four questions:

\noindent\textbf{Q1: Metric complementarity.}
Do full-pool retrieval metrics and target-over-confounder metrics reveal different ranking failures?

\noindent\textbf{Q2: Interface capacity.}
How much ranking performance is achieved by methods with no GraphRareBench supervision, prompted LLMs, and supervised graph-evidence rankers?

\noindent\textbf{Q3: Confounder mechanisms.}
Which types of graph-defined hard confounders remain difficult to distinguish from the target?

\noindent\textbf{Q4: Evidence and trace audit.}
What can fixed-feature interventions and tool-use traces reveal about the evidence used to produce a ranking?

\subsection{Experimental Protocol}

All full-pool methods rank the same fixed candidate set for each case. At test time, the target identity, hard-confounder labels, candidate-construction roles, graph-derived ranks and scores, and other evaluator-only fields are hidden unless they are explicitly included in the evaluated interface.

The gene-component-aware split contains 1,892 training cases, 236 validation cases, and 237 held-out test cases. On the test partition, each case contains a median of 69 full-pool candidates, including a median of five hard confounders. We report full-pool MRR, Hit@1, Hit@5, Hit@10, and case-averaged target-over-confounder accuracy. Analyses by confounder mechanism use pair-averaged target-over-confounder accuracy. Confidence intervals are estimated through case-level bootstrap resampling. Paired bootstrap resampling is used for method contrasts evaluated on the same cases.

LLM and agent results are obtained from one fixed protocol run. Accordingly, these results characterize behavior under the reported prompts, model versions, decoding settings, and candidate orders rather than variation across repeated model runs. Display-order analyses are performed on the outputs of these fixed runs.

\paragraph{Compared methods.}
Random ranking and shuffled-graph ranking serve as negative controls. HPO Similarity, Exomiser, and LIRICAL are phenotype-driven methods that use no GraphRareBench training labels. DeepSeek-V4-Flash \citep{deepseekai2026v4} is evaluated under three input conditions: coarsened HPO terms, source-support HPO terms, and compact candidate evidence cards. RAG2D, PPP, and LPP are trained on the GraphRareBench training and validation partitions using the same 21-feature interface. RAG2D provides a linear reference, whereas PPP and LPP provide nonlinear pairwise and listwise ranking models over the same 21-feature interface.

\subsection{Q1--Q2: Full Candidate-Pool Ranking}

\paragraph{Full-pool retrieval and hard-confounder ordering capture different failures.}
Table~\ref{tab:main-evidence} reports results on the held-out test partition. Random ranking yields an MRR of 0.070, while the shuffled-graph control yields 0.077. HPO Similarity achieves a relatively low full-pool MRR of 0.170 but ToC$_{\mathrm{case}}$ of 0.887. It therefore often ranks the target above the labeled hard confounders even when many additional candidates remain above the target in the full pool. This contrast shows that target retrieval and discrimination against the hard-confounder subset measure different aspects of ranking performance.

To examine failures that remain hidden by Hit@10, we compute the proportion of Hit@10 cases in which at least one hard confounder is still ranked above the target. Even among cases counted as Hit@10 successes, at least one graph-defined hard confounder remained above the target in 43.7\% of DeepSeek cases, 29.3\% of RAG2D cases, and 22.1\% of PPP cases. Thus, a target can satisfy the conventional Hit@10 criterion while remaining below a clinically plausible alternative.

\paragraph{Phenotype-driven tools.}
Among methods that use no GraphRareBench supervision, Exomiser and LIRICAL achieve MRRs of 0.322 and 0.502, respectively. Their native scores do not cover every candidate: Exomiser scores 74.6\% of candidates and 97.9\% of targets, while LIRICAL scores 80.2\% of candidates and 94.9\% of targets. Replacing the primary unscored-tail convention with tied average ranks changed MRR by less than \(10^{-4}\), indicating that the reported MRRs are insensitive to this specific tail-ranking convention.

\paragraph{LLM input conditions.}
DeepSeek-V4-Flash achieves an MRR of 0.247 when given the coarsened HPO query. Replacing the coarsened terms with the curator-derived source-support terms increases MRR to 0.313, corresponding to a paired improvement of 0.066 (95\% CI: 0.032--0.102). Because these source terms provide the original disease-level phenotype support from which the visible query was constructed, this condition represents a more informative upper-bound input rather than the standard benchmark setting.

Providing compact evidence cards changes DeepSeek's MRR from 0.247 to 0.219, while increasing its case-averaged target-over-confounder accuracy from 0.686 to 0.772. The paired changes are \(\Delta\mathrm{MRR}=-0.028\) (95\% CI: -0.069--0.012) and \(\Delta\mathrm{ToC}_{\mathrm{case}}=0.086\) (95\% CI: 0.039--0.133). In this fixed run, the evidence cards therefore improve ordering against hard confounders without improving the target's position in the full candidate pool. This result is specific to the reported evidence format and protocol; it does not establish a general effect of evidence augmentation or evaluate probability calibration.

\paragraph{Supervised graph-evidence rankers.}
Using GraphRareBench training and validation supervision, the linear RAG2D reranker achieves an MRR of 0.640, Hit@1 of 0.464, Hit@10 of 0.966, and \(\mathrm{ToC}_{\mathrm{case}}\) of 0.898. Its paired MRR difference is 0.138 relative to LIRICAL (95\% CI: 0.073--0.202), 0.393 relative to DeepSeek with coarsened HPO input (95\% CI: 0.337--0.448), and 0.421 relative to the DeepSeek evidence-card condition (95\% CI: 0.372--0.470). These comparisons show the level of performance attainable when the shared graph-evidence features are fitted using in-benchmark supervision. Because the compared methods differ in both supervision and information interface, the contrasts should be interpreted as interface-level performance comparisons rather than isolated estimates of model-architecture effects.

On the same 21-feature interface, the nonlinear PPP and LPP rankers achieve MRRs of 0.740 and 0.730, respectively. PPP and LPP achieve MRR point estimates that are 0.100 and 0.090 higher than RAG2D, respectively. Because the models differ in both functional form and ranking objective, these results suggest that higher-capacity ranking models can extract additional predictive signal from the shared feature interface.

\begin{table}[t]
\centering
\scriptsize
\setlength{\tabcolsep}{2.5pt}
\renewcommand{\arraystretch}{1.04}
\resizebox{\textwidth}{!}{%
\begin{tabular}{@{}p{0.19\textwidth}p{0.26\textwidth}p{0.12\textwidth}rrrrrl@{}}
\toprule
Method & Input interface & Supervision & MRR & Hit@1 & Hit@5 & Hit@10
& ToC$_{\mathrm{case}}$ & MRR 95\% CI \\
\midrule
\multicolumn{9}{c}{\textbf{Negative controls}} \\
\rowcolor{GRBControl}
Random Ranking & Candidates & None
& 0.070 & 0.013 & 0.059 & 0.173 & 0.532 & [0.056, 0.088] \\
\rowcolor{GRBControl}
Shuffled Graph & Permuted disease--HPO profiles & None
& 0.077 & 0.021 & 0.093 & 0.139 & 0.488 & [0.058, 0.098] \\
\midrule
\multicolumn{9}{c}{\textbf{Zero-label phenotype tools}} \\
\rowcolor{GRBPhenotype}
HPO Similarity & Coarsened HPO & None
& 0.170 & 0.072 & 0.219 & 0.384 & 0.887 & [0.139, 0.204] \\
\rowcolor{GRBPhenotype}
Exomiser & Coarsened HPO & External tool
& 0.322 & 0.160 & 0.489 & 0.684 & 0.748 & [0.282, 0.365] \\
\rowcolor{GRBPhenotype}
LIRICAL & Coarsened HPO & External tool
& 0.502 & 0.359 & 0.692 & 0.793 & 0.774 & [0.451, 0.552] \\
\midrule
\multicolumn{9}{c}{\textbf{Prompt-only LLM}} \\
\rowcolor{GRBLLM}
DeepSeek-V4-Flash & Coarsened HPO + candidates & Prompt
& 0.247 & 0.110 & 0.376 & 0.570 & 0.686 & [0.211, 0.286] \\
\rowcolor{GRBLLM}
DeepSeek-V4-Flash & Source-support HPO + candidates & Prompt
& 0.313 & 0.160 & 0.477 & 0.654 & 0.711 & [0.272, 0.357] \\
\rowcolor{GRBLLM}
DeepSeek-V4-Flash & Coarsened HPO + evidence cards & Prompt
& 0.219 & 0.072 & 0.346 & 0.549 & 0.772 & [0.187, 0.252] \\
\midrule
\multicolumn{9}{c}{\textbf{Supervised train/dev graph-evidence rankers}} \\
\rowcolor{GRBGraph}
RAG2D linear reranker & 21 graph-evidence features & Train/dev
& 0.640 & 0.464 & 0.899 & 0.966 & 0.898 & [0.595, 0.686] \\
\rowcolor{GRBProbe}
PPP nonlinear pairwise & 21 graph-evidence features & Train/dev
& \textbf{0.740} & \textbf{0.595} & \textbf{0.928} & \textbf{0.975}
& \textbf{0.916} & \textbf{[0.697, 0.782]} \\
\rowcolor{GRBProbe}
LPP nonlinear listwise & 21 graph-evidence features & Train/dev
& 0.730 & 0.586 & 0.903 & 0.970 & 0.903 & [0.687, 0.773] \\
\bottomrule
\end{tabular}
}
\caption{Full candidate-pool ranking results on the held-out test partition.
All methods rank the same released \(C_i^{\mathrm{full}}\). MRR and
Hit@\(k\) score target recovery, whereas ToC\(_{\mathrm{case}}\) scores
target-over-hard-confounder ordering. RAG2D, PPP, and LPP share the same
21-feature interface and use GraphRareBench train/dev supervision; cross-block
comparisons quantify performance across different information regimes.}
\label{tab:main-evidence}
\end{table}
\subsection{Q3: Reason-Sliced Confounder Analysis}

\begin{figure}[t]
\centering
\includegraphics[width=0.7\textwidth]{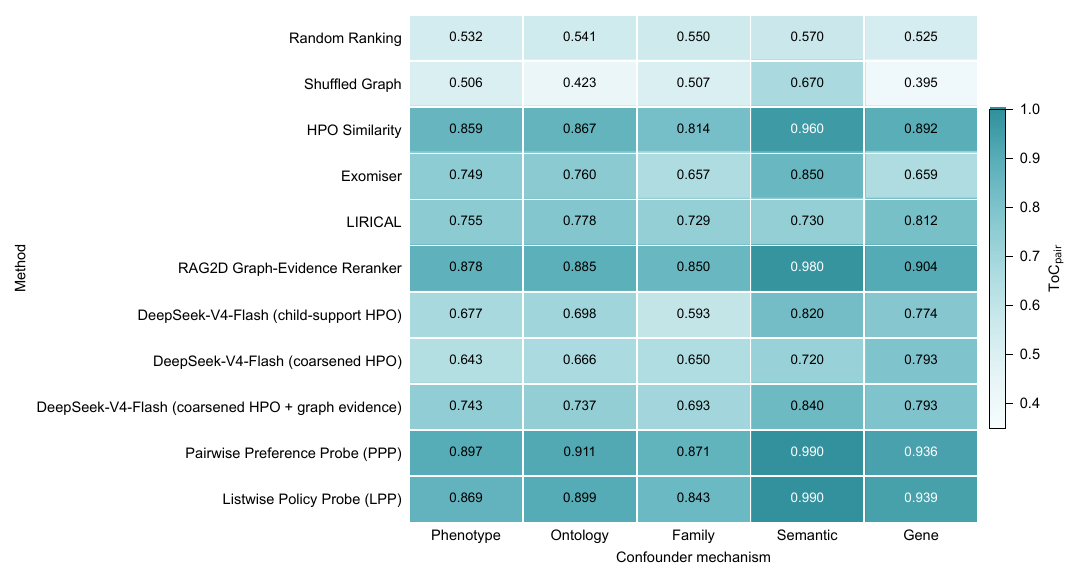}
\caption{Reason-sliced target-over-confounder heatmap on the test partition.
Entries are ToC$_{\mathrm{pair}}$, the fraction of target--confounder pairs
where the target disease is ranked above the hard confounder. Pairs with
multiple reason labels contribute to each matching column. Slice case/pair
counts are phenotype 237/474, ontology 186/338, family 98/140, semantic
94/100, and gene 148/314.}
\label{fig:reason-sliced-toc}
\end{figure}

Figure~\ref{fig:reason-sliced-toc} reports mechanism-specific \(\mathrm{ToC}_{\mathrm{pair}}\); multi-label pairs contribute to every applicable, and therefore overlapping, mechanism slice. The supervised rankers perform particularly well on semantic-neighbor confounders: RAG2D achieves a \(\mathrm{ToC}_{\mathrm{pair}}\) of 0.980, while PPP and LPP both reach 0.990. Performance is lower for disease-family confounders, with scores of 0.850 for RAG2D, 0.871 for PPP, and 0.843 for LPP. Gene-context confounders also reveal substantial differences between methods: Exomiser achieves 0.659, compared with 0.936 and 0.939 for PPP and LPP, respectively. These results show that strong aggregate ranking performance does not imply uniform discrimination across confounder types. Across the supervised rankers, error rates were consistently higher for disease-family and gene-context confounders than for semantic-neighbor confounders. 2,530 of 18,093 pairs (14.0\%) carried more than one mechanism label. Confidence intervals for the reason-sliced results use case-clustered resampling so that multiple confounders and overlapping labels from the same case are not treated as independent observations.

\subsection{Q4: Evidence-Channel and Agent-Trace Audit}

GraphRareBench audits evidence through two complementary views. First, we
apply score-time feature isolation to the frozen PPP checkpoint while keeping the evaluation cases, training, and model selection fixed. Recomputed scores reproduce the stored rankings, with a maximum difference of \(1.53\times10^{-5}\) and no rank mismatches. Because PPP is nonlinear, Table~\ref{tab:ppp-channel-isolation} measures retained-set sufficiency rather than independent feature contribution or necessity. With all 21 features, MRR is 0.740; the case-specific and disease-profile/context sets retain MRRs of 0.510 and 0.547, whereas the aggregate graph-score and strict non-query sets yield 0.124 and 0.024. Aggregate graph scores retain some target-over-confounder separation, while strict non-query features preserve little ranking signal. Second, the tool-mediated audit records the disease profiles and query-conditioned evidence requested before ranking.

\begin{table}[t]
\centering
\normalsize
\setlength{\tabcolsep}{3pt}
\renewcommand{\arraystretch}{1.08}
\begin{tabular}{@{}lrrrrrr@{}}
\toprule
PPP score-time variant & Feat. & MRR & Hit@1 & Hit@5 & Hit@10 & ToC$_{\mathrm{case}}$ \\
\midrule
Full PPP checkpoint
& 21 & 0.740 & 0.595 & 0.928 & 0.975 & 0.916 \\
Case-specific evidence only
& 8 & 0.510 & 0.316 & 0.776 & 0.873 & 0.851 \\
Disease-profile/context only
& 11 & 0.547 & 0.380 & 0.793 & 0.907 & 0.846 \\
Aggregate graph scores only
& 2 & 0.124 & 0.008 & 0.169 & 0.371 & 0.719 \\
Strict non-query profile/centrality only
& 6 & 0.024 & 0.000 & 0.000 & 0.000 & 0.565 \\
\bottomrule
\end{tabular}
\caption{Score-time feature-channel isolation for the frozen PPP checkpoint. Each row retains only the indicated feature group while keeping training and model selection fixed. Results reflect checkpoint sensitivity rather than independent feature contributions.}
\label{tab:ppp-channel-isolation}
\end{table}
We next evaluate tool-using agents on the hard-confounder pool. For each case, the agent receives a coarsened phenotype query and a shuffled list of candidate disease names and IDs. It may make four tool calls before returning a complete ranking of all candidates. Two tools are available: a disease-profile tool that retrieves general information about selected diseases and a query-conditioned evidence tool that retrieves evidence linking selected candidates to the patient phenotype query. Each call can include up to three candidate diseases.

We evaluate agents instantiated with DeepSeek-V4-Flash and Agents-A1 \citep{bai2026agentsa1}. Deterministic checks verify that each run respects the permitted information boundary, makes only valid tool calls, and returns a complete candidate ranking. An independent LLM auditor additionally reviews compliance with the tool-use protocol. Ranking metrics are computed directly from the final candidate order.

As shown in Table~\ref{tab:hard-pool-tool-agents}, the Agents-A1-based agent achieves an MRR of 0.746, compared with 0.718 for the DeepSeek-based agent. The paired MRR difference is 0.029, with a 95\% confidence interval of -0.010 to 0.068 and \(p=0.155\). Because the confidence interval includes zero, the observed difference in final ranking performance is not statistically significant.

The two agents nevertheless exhibit large descriptive differences in tool-use patterns in this fixed run. The DeepSeek-based agent makes an average of 2.11 query-conditioned evidence calls per case, whereas the Agents-A1-based agent makes 0.49. It also switches more frequently between the disease-profile and query-conditioned evidence tools, averaging 1.65 switches per case compared with 0.58 for the Agents-A1-based agent. Correspondingly, the DeepSeek-based agent retrieves evidence for a larger proportion of both target diseases and hard confounders. Its target-evidence coverage is higher by 0.561 (95\% CI: 0.494--0.624), and its hard-confounder evidence coverage is higher by 0.609 (95\% CI: 0.570--0.648). In this fixed protocol run, an MRR difference of 0.029 co-occurred with substantially different evidence-seeking patterns, including a target-evidence coverage difference of 0.561 and a hard-confounder evidence coverage difference of 0.609.

\begin{table}[t]
\centering
\normalsize
\setlength{\tabcolsep}{3pt}
\renewcommand{\arraystretch}{1.08}
\begin{tabular}{@{}lrr@{}}
\toprule
Metric & DeepSeek-V4-Flash & Agents-A1 \\
\midrule
MRR & 0.718 & 0.746 \\
Hit@1 & 0.540 & 0.582 \\
ToC$_{\mathrm{case}}$ & 0.817 & 0.833 \\
Profile calls/case & 1.89 & 3.51 \\
Evidence calls/case & 2.11 & 0.49 \\
Candidates observed/call & 2.58 & 2.34 \\
Unique candidates/case & 5.79 & 5.54 \\
Tool switches/case & 1.65 & 0.58 \\
Target evidence coverage & 0.937 & 0.376 \\
Hard-confounder evidence coverage & 0.863 & 0.254 \\
\bottomrule
\end{tabular}
\caption{Tool-mediated hard-pool audit. Call and switch counts are averaged per case; candidate observations are averaged per tool call. Evidence coverage measures the proportion of targets or hard confounders for which query-conditioned evidence was retrieved.}
\label{tab:hard-pool-tool-agents}
\end{table}
\section{Discussion}

GraphRareBench provides a controlled basis for evaluating modular diagnostic
agents. A planning component can select candidates and evidence sources for
further investigation, while specialized executors perform phenotype
matching, graph traversal, gene-context retrieval, or source verification.
Because candidate-level actions are aligned with hidden confounder annotations
and source-linked evidence, the benchmark can separate gains arising from
search policies, evidence executors, and allocation of a limited tool budget.

The fixed candidate pools also support controlled perturbations of phenotype
granularity, candidate order, evidence availability, graph relations, source
diversity, and tool-call budgets. Such interventions can distinguish failures
caused by unavailable evidence from failures of evidence retrieval or
integration. An open-world extension could first retrieve diseases from a
larger knowledge base and then evaluate whether the target is correctly
ordered against the hard alternatives in the retrieved set. Longitudinal
phenotypes, uncertain or negated findings, laboratory measurements, and
variant-level evidence provide further extensions.

Finally, the frozen candidate pools, evidence views, confounder mechanisms, and trace checks make GraphRareBench suitable as a regression suite. Updates to models, retrievers, or knowledge bases can be evaluated for aggregate gains and for localized regressions across confounder mechanisms, phenotype abstraction, query-independent shortcuts, and evidence-acquisition behavior.
\section{Limitations}

GraphRareBench evaluates ontology-derived cases within fixed candidate pools and therefore does not capture the full complexity of open-world clinical diagnosis, including noisy patient narratives, longitudinal phenotype evolution, uncertain or negated findings, and variant-level interpretation. Its hard confounders are defined through reproducible graph and semantic relations; although these relations provide controlled and challenging alternatives, they may not cover every distinction encountered in clinical practice. 
\section{Conclusion}

We introduced GraphRareBench, a provenance-preserving benchmark for phenotype-driven rare-disease ranking. GraphRareBench combines coarsened phenotype queries, fixed candidate pools, graph-defined hard confounders, source-linked evidence, and hidden evaluator annotations within a unified evaluation framework. Its full-pool and target-over-confounder metrics reveal complementary ranking failures that conventional top-\(k\) evaluation can obscure. Feature-channel interventions and tool-mediated traces further distinguish final ranking performance from the evidence accessed to produce it. By jointly evaluating target retrieval, discrimination against plausible alternatives, and observable evidence-seeking behavior, GraphRareBench provides a reproducible test bed for developing and auditing evidence-aware rare-disease diagnostic systems.

\bibliography{references_revised}

\end{document}